\documentclass[runningheads]{cl2emult}

\usepackage{makeidx}  
\usepackage{graphicx} 
\usepackage{subeqnar} 
\usepackage{multicol} 
\usepackage{cropmark} 
\usepackage{eso}      
\makeindex            

\begin{document}
\title*{Data Analysis for the Microwave Anisotropy Probe (MAP) Mission}
\toctitle{Data Analysis for the Microwave Anisotropy Probe
\protect\newline (MAP) Mission}
%
%
\titlerunning{Data Analysis for MAP}
%
\author{Gary Hinshaw\inst{1}}
\authorrunning{Gary Hinshaw}
%
%
\institute{Code 685, Goddard Space Flight Center, Greenbelt MD 20771, USA, \\
           for the MAP Science Team}

\maketitle              

\begin{abstract}
We present an overview of the upcoming Microwave Anisotropy Probe (MAP)
mission, with an emphasis on those aspects of the mission that simplify the
data analysis.  The method used to make sky maps from the differential 
temperature data is reviewed and we present some of the noise properties
expected from these maps.  An overview of the method we plan to use to mine the
angular power spectrum from the mega-pixel sky maps closes the paper.
\end{abstract}

\section{Mission Overview}

In 1992 NASA's Cosmic Background Explorer (COBE) satellite made a full sky map 
of the cosmic microwave background (CMB) temperature with 7$^\circ$ 
resolution, uncorrelated pixel noise, minimal systematic errors, and accurate
calibration, from which CMB temperature anisotropy was first discovered
\cite{smoot92},  \cite{bennett92}, \cite{wright92}, \cite{bennett96}.  The
purpose of the MAP mission is to re-map the anisotropy over the full sky with
more than 30 times the angular resolution ($\sim 0.23^{\circ}$ FWHM) and more
than 35 times the sensitivity  ($\sim 20$ $\mu$K per 0.3$^{\circ}$ pixel) of
COBE, but with the same level of quality control as was possible with COBE.
With this data, MAP will measure the physical interactions of the photon-baryon
fluid (sound waves) in the early universe and thereby test models of structure
formation, the geometry of the universe, and inflation.

In the years since COBE, a host of ground-based and balloon-borne experiments
have detected and characterized fluctuations at smaller angular scales, most
recently the experiments TOCO \cite{toco}, BOOMERanG \cite{boomldb} and MAXIMA
\cite{maxima}. However, because of their proximity to the Earth and its 
atmosphere, none of the ground or balloon-based experiments enjoy the extent 
of systematic error rejection or calibration accuracy that was possible with 
COBE. Moreover, many of these experiments have significantly correlated noise 
that places severe demands on the data analysis.  The COBE data still serves 
as a benchmark for the field and many aspects of the COBE mission have 
influenced the design of the MAP mission.  The need to minimize the level of 
systematic errors has been the major driver of the MAP design and has led to 
the following high level design features:

\begin{itemize}
\item
{\bf A highly symmetric differential design}: MAP is a differential
experiment based on pseudo-correlation microwave  radiometers that employ
phase-matched HEMT amplifiers.  The instrument measures temperature differences
between two  points $\sim$ 141$^{\circ}$ apart on the sky.  By measuring
temperature differences, rather than absolute temperatures, many spurious
signals will be common-mode and thus cancel upon differencing.  Also, by
employing a pseudo-correlation design with a fast chopping frequency between
the two sky inputs, 1/f noise that arises from the HEMTs can be chopped out. 
The resulting power spectrum of the radiometer noise is very nearly white (see
\S 3).

\item
{\bf Multi-frequency}: There are five frequency bands from 22-90 GHz that will
allow emission from the Galaxy and other non-cosmological sources to be modeled
and removed based on their frequency dependence.  In the lowest frequency
bands, MAP will probe the high frequency tail of radio emission from our Galaxy
and provide valuable data on the enigmatic microwave foreground emission that
correlates with thermal dust emission, but has a much different spectrum.  See 
\cite{kogut} for a recent summary of evidence for this foreground.

\item
{\bf Stability}: MAP will observe from a Lissajous orbit about the L2 Lagrange
point 1.5 million km from Earth. The L2 point offers an exceptionally stable
environment and an unobstructed view of deep space, with the Sun, Earth, and
Moon always in shadow behind MAP's Sun shield.  MAP's large distance from Earth
protects it from near-Earth emission and other disturbances.  While observing
at L2, MAP's Sun shield and solar panels maintain a fixed angle with respect to
the Sun  to provide exceptional thermal and power stability.

\item
{\bf Low beam sidelobe levels}: The MAP optical system was designed with the
foremost goal of providing adequate angular resolution along the line of sight
while at the same time rejecting stray light from other directions.  For
example, the largest instantaneous signal due to radiation from the galactic plane
spilling into a sidelobe is expected to be less than 2 $\mu$K at 90 GHz.
\end{itemize}

\begin{figure}
\centering
\includegraphics*[width=0.8\hsize,keepaspectratio]{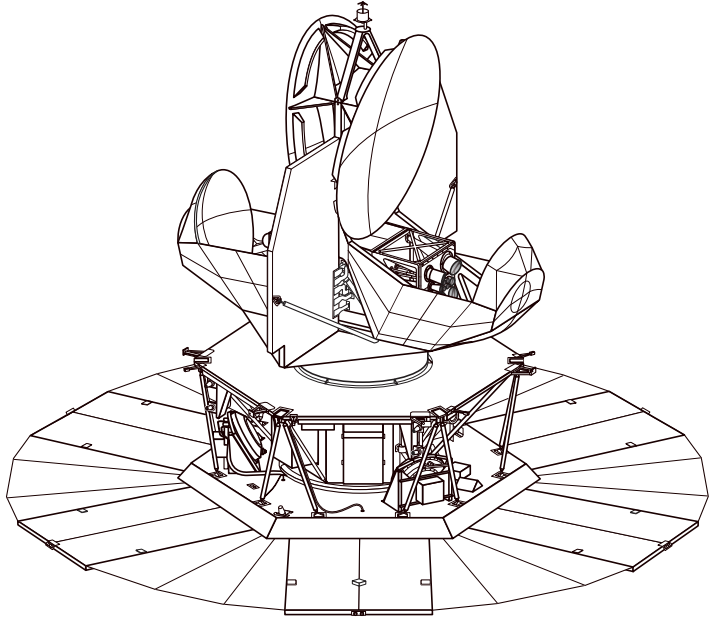}
\caption{\small An overview of the MAP satellite.}
\label{map_view}
\end{figure}

An overview of the MAP satellite is shown in Fig.~\ref{map_view}.  The major
visible features of MAP include the back-to-back telescope optics with 1.4 
$\times$ 1.6 m primary mirrors and 1 m secondary mirrors, the passive thermal
radiators which cool the HEMT amplifiers to $<$100 K, the hexagonal structure
housing the spacecraft service modules, and the large solar panel array/Sun
shield which keeps the instrument in full shade.  MAP weighs a total of 830 kg
and stands $\sim$4 m tall.  It will be launched in 2001 aboard a Delta 7425-10
expendable launch vehicle from the NASA Kennedy Space Center Eastern Test
Range.

\section{Scan Strategy}

The MAP scan strategy plays an important role in systematic error rejection.  
It was designed with the following goals in mind:
\begin{itemize}
\item
Scan a large fraction of the sky as rapidly as possible, consistent with 
reasonable requirements on the controlling hardware and the telemetry data rate. 
\item
Scan each sky pixel through as many azimuthal angles as possible for the 
reasons listed below.
\item
Observe a given pixel on as many different time scales as possible.
\item
Maintain the instrument in continuous shadow for optimal passive cooling and 
avoidance of stray signals from the Sun, Earth, and Moon.
\item
Maintain a constant angle between the Sun and the plane of the solar panels 
for thermal and power stability.
\end{itemize}
The strategy that was ultimately adopted combines a ``fast'' spin about the
spacecraft symmetry axis with a slow precession about the Sun-MAP line (which
is always within 0.1$^{\circ}$ of the Sun-Earth line at L2).  Since each
telescope line of sight is $\sim 70^{\circ}$ off the symmetry axis, the path
swept out on the sky by a given line of sight resembles a spirograph pattern
that reaches from the north to south ecliptic poles.  As MAP orbits the Sun,
this pattern revolves around the sky so that full sky coverage is first
achieved after 6 months of observing at L2.

\begin{figure}
\centering
\includegraphics*[width=0.8\hsize,keepaspectratio]{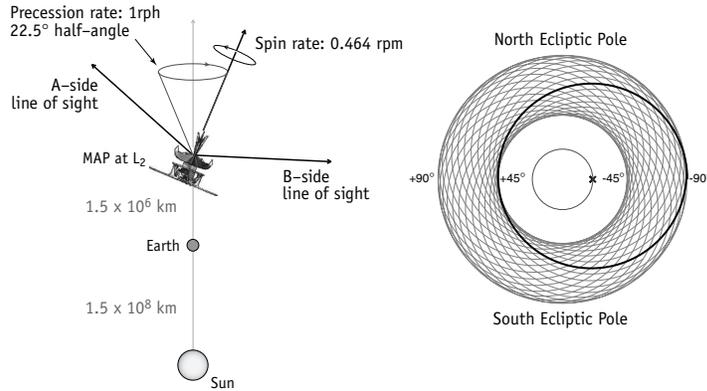}
\caption{\small A schematic of the MAP scan strategy from L2. The satellite 
covers approximately 35\% of the sky each day.  The scan pattern on the right
depicts the motion of a single line of sight over a 1 hour precession period
(lighter spirograph pattern), and a $\sim$2 minute spin period (darker single
circle). The inner circle depicts the path of the spin axis during one
precession.}
\label{scan_strategy}
\end{figure}

The MAP scan strategy achieves a reasonable level of azimuthal coverage in each
sky pixel.  For example, a pixel in the ecliptic equator is observed over 
$\sim$30\% of the possible angles of attack; a pixel at the cusp of the annular
coverage at $\sim 45^\circ$ ecliptic latitude is observed over about 70\% of
possible angles of attack; and a pixel near the ecliptic poles is observed
from 100\% of the possible azimuthal orientations.  A large azimuthal coverage 
provides numerous desirable features in the data:
\begin{itemize}
\item
Helps to produce a stable sky map solution.
\item
Produces small pixel-pixel covariance at the beam separation scale.
\item
Minimizes striping due to any residual 1/f noise in the differential data.
\item
Maximizes polarization sensitivity.
\item
Maximizes azimuthal symmetry of the beam response on the sky.
\end{itemize}
From this standpoint, the MAP strategy is not as complete as  COBE's, which
achieved nearly 100\% azimuthal coverage in all pixels.  However, in order to
achieve such completeness, the spacecraft spin axis must ultimately point to
every pixel on the sky which is less desirable from the standpoint of
systematic error avoidance.  The MAP strategy achieves reasonable azimuthal
coverage consistent with strong systematic error constraints.  Extensive
simulations of the map making and power spectrum estimation procedures have
shown that the strategy is more than adequate to meet MAP's scientific goals.


\section{Map Making with Differential Data}

Sky maps, ${\bf t}$, are obtained from the differential data, ${\bf d}$, by
linear least squares fitting.  The raw data takes the form ${\bf d} = 
{\bf A}\cdot{\bf t}$ where ${\bf A}$ is the scan matrix of the experiment 
which has $N_{pix}$ columns and $N_{obs}$ rows, where $N_{pix}$ is the number 
of sky map pixels and $N_{obs}$ is the total number of differential
observations.  

\begin{figure}
\centering
\includegraphics*[width=0.9\hsize,keepaspectratio]{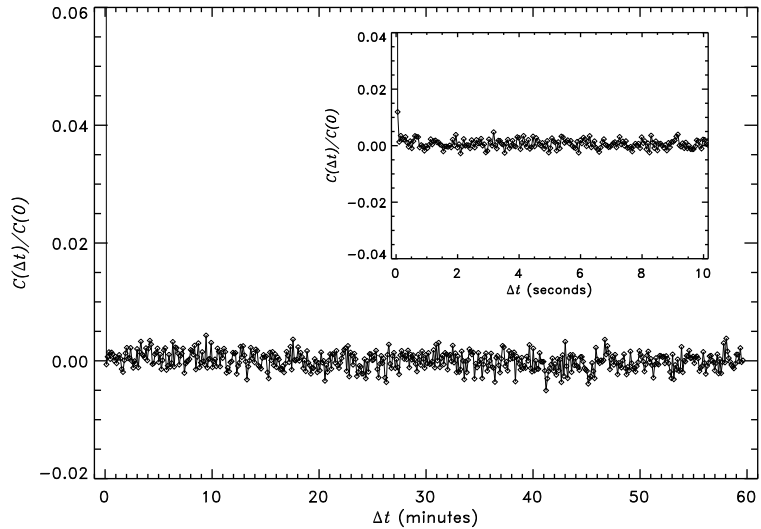}
\caption{\small The auto-correlation function (denoted ${\bf T}$ in the text) 
of 8 hours of noise from one of the MAP 90 GHz radiometers, taken during a cold
instrument test in March 2000.  The instrument was operating at expected flight
temperature while  observing temperature controlled targets over the flight
feeds.  The only  processing applied to the raw data was to subtract a baseline
with power on  periods of greater than ~1 hour (a Legendre  polynomial of order
8 fit  over the 8 hour period of data).  This approximates the procedure that
will be applied to the flight data.  Note the point at lag 1 (inset) has a
covariance of 1.2\% of the raw noise due to a low-pass filter applied just 
prior to the a/d conversion.  There is no other significant covariance in the 
noise.}
\label{noise}
\end{figure}

Each row (observation) of length $N_{pix}$ contains a $+1$ in the element 
(pixel) observed by the $+$ horn during that observation, and a $-1$ in the 
element observed by the $-$ horn.  The normal equations for the sky map 
solution are
\begin{equation}
({\bf A}^T\cdot{\bf T}^{-1}\cdot{\bf A})\cdot{\bf t} 
= {\bf A}^T\cdot{\bf T}^{-1}\cdot{\bf d}
\end{equation}
where ${\bf T}$ is the covariance of the instrument noise in the time domain. 
For MAP's differential radiometers this is well approximated by stationary, 
white noise, $T_{tt'} = \langle n_t n_{t'} \rangle \approx \sigma_0^2 
\delta_{tt'}$.  A sample of the MAP noise covariance is shown in
Fig.~\ref{noise} from test data taken with the flight instrument. The
whiteness of the noise greatly simplifies the subsequent data processing and
analysis for two reasons 1) the normal equations for the sky map are much
easier to solve than they would be with significant 1/f noise, 2) the
pixel-pixel noise covariance in the final sky maps is very small and completely
negligible for most applications.

With stationary white noise, the maximum likelihood sky map solution reduces to 
\begin{equation}
{\bf t} = ({\bf A}^T \cdot {\bf A})^{-1}
\cdot ({\bf A}^T \cdot {\bf d})
\label{map_soln}
\end{equation}
while the pixel-pixel noise covariance is
\begin{equation}
{\bf N} = \sigma_0^2\,({\bf A}^T \cdot {\bf A})^{-1}.
\end{equation}
Formally, the sky map solution (\ref{map_soln}) requires inverting an 
$N_{pix} \times N_{pix}$ matrix ${\bf M} \equiv ({\bf A}^T \cdot {\bf A})$ 
which has the form
\begin{equation}
{\bf M} = \left( \begin{array}{rrrrrrrr}
N_0    &  0     &  0     & \cdots &-N_{0i} & 0      & 0      & \cdots \\
0      &  N_1   &  0     & \cdots & 0      &-N_{1j} & 0      & \cdots \\
0      &  0     &  N_2   & \cdots & 0      & 0      &-N_{2k} & \cdots \\
\vdots & \vdots & \vdots & \vdots & \vdots & \vdots & \vdots & \ddots \\
-N_{0i}&  0     &  0     & \cdots & N_i    & 0      & 0      & \cdots \\
0      & -N_{1j}&  0     & \cdots & 0      & N_j    & 0      & \cdots \\
0      &  0     & -N_{2k}& \cdots & 0      & 0      & N_k    & \cdots
\end{array} \right)
\end{equation}
where the $i$th diagonal element is the number of times pixel $i$ was observed  by
either side of the instrument, and the $ij$th off-diagonal element is the 
number of times pixels $i$ and $j$ were observed simultaneously.  Because of 
the fixed separation between the two instrument beams, only pairs of pixels 
separated by a fixed distance can be simultaneously observed, thus this matrix 
is very sparse.  Furthermore, because the scan strategy generates ample 
azimuthal coverage, we have $N_{ij} \ll N_i$ so ${\bf M}$ is diagonally
dominant.  [Formally ${\bf M}$ is singular because the sum of each of its
rows or columns is zero, due to the fact that the time-series data is 
differential.  Thus the sky map solution and its pixel-pixel covariance are 
undetermined up to a constant since ${\bf M} \cdot {\bf C} = 0$ for any 
constant vector or matrix ${\bf C}$.  In practice this mode is readily 
projected out of the data in any actual inversion scheme.]

The form of ${\bf M}$ suggests an iterative approach to solving for the map 
that employs a diagonal pre-conditioner as an approximate inverse 
\cite{num_rec}.  Define
\begin{equation}
\tilde{\bf M}^{-1} = 
\mbox{diag}(\frac{1}{N_0},\frac{1}{N_1},\frac{1}{N_2},\ldots)
\end{equation}
and assume an approximate initial map solution ${\bf t}_0 = {\bf t} + 
\delta{\bf t}$ such that
\begin{equation}
{\bf M} \cdot {\bf t}_0 
= {\bf M} \cdot ({\bf t} + \delta {\bf t}) 
= {\bf v} + \delta{\bf v}
\end{equation}
where ${\bf v} \equiv {\bf A}^T \cdot {\bf d}$.  This gives an expression for 
$\delta {\bf t}$ in terms of known quantities
\begin{equation}
{\bf M} \cdot \delta {\bf t} = {\bf M} \cdot {\bf t}_0 - {\bf v}
\end{equation}
We can use the above pre-conditioner to obtain an iterative estimate of the map
correction $\delta {\bf t}_1$, and hence an improved solution ${\bf t}_1$
\begin{equation}
{\bf t}_1 = {\bf t}_0 - \delta {\bf t}_1
          = {\bf t}_0 - \tilde{{\bf M}}^{-1} \cdot ({\bf M} \cdot {\bf t}_0 - {\bf v}).
\label{map_it}
\end{equation}
As was noted in \cite{wright96}, an efficient way to evaluate the right hand
side of (\ref{map_it}) is to group the operations as follows
\begin{equation}
{\bf t}_{n+1} = (\tilde{{\bf M}}^{-1} \cdot {\bf A}^T) \cdot
              ({\bf d} - {\bf A} \cdot {\bf t}_n) + {\bf t}_n.
\label{map_wright}
\end{equation}
The interpretation of (\ref{map_wright}) is that for each pixel the new
sky map temperature is equal the average of all differential observations of
that pixel, accounting for the sign of the observing horn, corrected by an
estimate of the signal in the pair horn, based on the previous sky map
iteration.  The expression in (\ref{map_wright}) can be efficiently evaluated
because the sums can be accumulated by reading through the time-series data,
read from disk, and accumulating data into arrays of length $N_{pix}$.  It is
never necessary to store or invert and $N_{pix} \times N_{pix}$ matrix.

\begin{figure}
\centering
\includegraphics*[width=0.9\hsize,keepaspectratio]{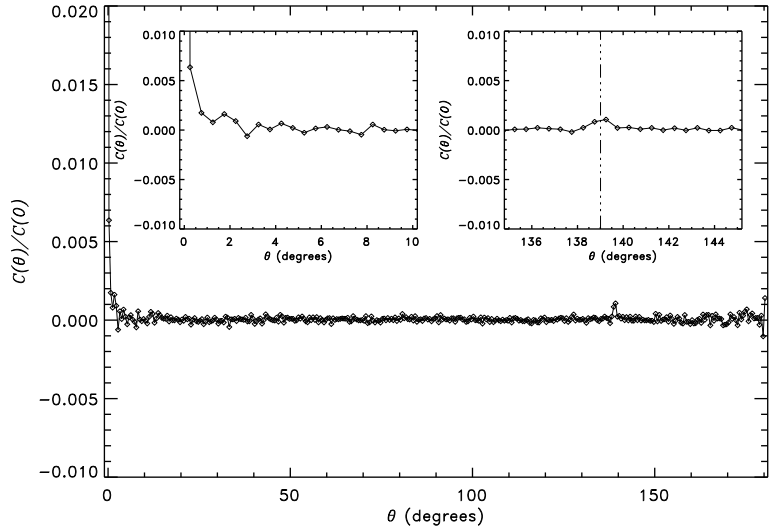}
\caption{\small 2-point angular correlation function of a simulated MAP noise
map.  The function was computed from a residual sky map that was obtained by
subtracting an input sky map from a solution of the map-making algorithm.  The
noise used in the simulated differential data was statistically identical to the
actual radiometer test data shown in Fig.~\ref{noise}.  Note the small
($<$1\%) nearest-neighbor covariance due to the nearest-neighbor covariance in
the time-series data.  There is also a small ($\sim$0.1\%) covariance between
pixels separated by the beam spacing, indicated by the dashed line in the inset
panel.}
\label{map_2pt}
\end{figure}

The performance of this algorithm has been tested extensively with simulations
that capture differential data with realistic instrument noise observed with
the MAP scan strategy.  The solution is seen to converge to the input in less
than 50 iterations, starting from an initial guess with no anisotropy.  In 
cases where the instrument noise has been suppressed for study purposes,
peak-peak features in the residual sky map (output $-$ input) are less than 0.1
$\mu$K.  In cases where realistic noise is included, the performance of the
algorithm can be measured by the 2-point angular correlation function of the
residual (noise) map.  Ideally, this would  be everywhere consistent with zero
except for the bin at zero angular separation which should equal the variance
of the map. Fig.~\ref{map_2pt} shows the 2-point function obtained from a noise
map that was obtained from the MAP algorithm.  To a very good approximation,
the noise in the sky map is uncorrelated from pixel to pixel.

\section{Power Spectrum Estimation from Full Sky Maps}

The most fundamental statistic to measure from the observed sky map is its
angular power spectrum, $C_{\ell}$, which measures the variance of the
fluctuations over a range of angular scales.  If the temperature fluctuations 
are gaussian, and the {\it a priori} probability of a given set of cosmological
parameters is uniform, then the power spectrum may be estimated by maximizing
the multi-variate gaussian likelihood function
\begin{equation}
{\cal L}(C_{\ell}|{\bf m}) = \frac{\exp(-\frac{1}{2}
{\bf m}^T \cdot {\bf C}^{-1} \cdot {\bf m})}
{\sqrt{\det {\bf C}}}
\end{equation}
where ${\bf m}$ is a data vector (see below) and ${\bf C}$ is the covariance
matrix of the data which has contributions from both the signal and the
instrument noise,  ${\bf C} = {\bf S} + {\bf N}$.  We can work in whatever
basis is most convenient, in the pixel basis the data are the sky map pixel
temperatures while in the spherical harmonic basis the data are the $a_{\ell m}$
coefficients of the map.  In the former basis the noise covariance is 
nearly diagonal, while in the latter, the signal covariance is
$$
\begin{array}{lllp{5mm}lll}
 & & \mbox{Pixel basis:} & & & & \mbox{Spherical harmonic basis:} \\[2mm]
{\bf m} & \rightarrow & T_i & & {\bf m} & \rightarrow & a_{\ell m} \\[2mm]
{\bf S} & \rightarrow & \sum_{\ell} \frac{(2\ell + 1)}{4\pi}C_{\ell}P_{\ell}(\cos\theta_{ij}) 
& & {\bf S} & \rightarrow & {\rm diag}(C_2,C_2,\ldots C_3,C_3,\ldots) \\[2mm]
{\bf N} & \rightarrow & \sigma_i^2 \delta_{ij} & & {\bf N} & \rightarrow & 
N_{(\ell m)(\ell m)^{\prime}} \mbox{(see below)}.
\end{array}
$$

In the case of MAP, the length of the data vector, $N_{data}$, is of order  1
million, so it is necessary to find methods for evaluating ${\cal L}$ that do
not  require a full inversion of the covariance matrix ${\bf C}$, which
requires  $O(N_{data}^3)$ operations.  As with the map-making, our approach is
fundamentally an iterative one that exploits the ability to find an approximate
inverse $\tilde{\bf C}^{-1}$.  The most important feature in the data that
makes this possible is the fact that we cover the full sky and that the galaxy
cut we impose on the data is predominantly azimuthally symmetric in galactic
coordinates.  Of secondary importance for this pre-conditioner is that fact 
that our noise per pixel is not too strongly varying across the sky.  We
discuss the pre-conditioner in more detail below.

We maximize the likelihood by solving
\begin{equation}
\begin{array}{lp{5mm}l}
\frac{\partial f}{\partial C_{\ell}} = 0
 = {\bf m}^T\cdot{\bf C}^{-1}\cdot{\bf P}^{\ell}\cdot{\bf C}^{-1}\cdot{\bf m} 
 + {\rm tr}({\bf C}^{-1}\cdot{\bf P}^{\ell}) & & 
 f \equiv -2\ln {\cal L} \\[2mm]
 & & {\bf P}^{\ell} \equiv \frac{\partial {\bf C}}{\partial C_{\ell}}
\end{array}
\end{equation}
using a Newton-Raphson approach to root finding that generates an iterative
estimate of the angular power spectrum at each step
\begin{equation}
C_{\ell}^{(n+1)} = C_{\ell}^{(n)} - \frac{1}{2}\sum_{\ell^{\prime}} 
F_{\ell\ell^{\prime}} \frac{\partial f}{\partial C_{\ell}}
\label{nr_cl}
\end{equation}
where $F_{\ell\ell^{\prime}}$ is the Fisher matrix
\begin{equation}
F_{\ell\ell^{\prime}} = 
\langle 
-\left(\frac{\partial^2}{\partial C_{\ell}\partial C_{\ell^{\prime}}}\right)
\ln {\cal L} \rangle
= \frac{1}{2}{\rm tr}({\bf C}^{-1}\cdot{\bf P}^{\ell}\cdot{\bf C}^{-1}\cdot{\bf P}^{\ell^{\prime}}).
\end{equation}
In order to implement the solution in (\ref{nr_cl}) we need to be able to 
evaluate the following components of $\sum_{\ell^{\prime}} F_{\ell\ell^{\prime}}
\frac{\partial f}{\partial C_{\ell}}$ quickly
\begin{equation}
\begin{array}{l}
{\bf m}^T\cdot{\bf C}^{-1}\cdot{\bf P}^{\ell}\cdot{\bf C}^{-1}\cdot{\bf m} \\
{\rm tr}({\bf C}^{-1}\cdot{\bf P}^{\ell}) \\
{\rm tr}({\bf C}^{-1}\cdot{\bf P}^{\ell}\cdot{\bf C}^{-1}\cdot{\bf P}^{\ell^{\prime}}).
\end{array}
\end{equation}
We use the spherical harmonic basis in which the data vector consists of the 
$a_{\ell m}$ coefficients of the map obtained by least squares fitting on the 
cut sky.  The signal covariance is diagonal in this basis, while the noise 
matrix is obtained from the normal equations for the
$a_{\ell m}$ fit
\begin{equation}
\sum_{(\ell m)^{\prime}} N^{-1}_{(\ell m)(\ell m)^{\prime}} a_{(\ell m)^{\prime}}
 =  y_{(\ell m)}
\end{equation}
where
\begin{equation}
\begin{array}{lp{5mm}l}
N^{-1}_{(\ell m)(\ell m)^{\prime}} \equiv \sum_i 
\frac{Y_{(\ell m)}(\hat{n}_i)Y_{(\ell m)^{\prime}}(\hat{n}_i)}{\sigma_i^2}
& &
y_{(\ell m)} \equiv \sum_i \frac{T_i Y_{(\ell m)}(\hat{n}_i)}{\sigma_i^2}.
\label{Ny_def}
\end{array}
\end{equation}
The sums are over the uncut pixels in the sky map, and we have used 
the fact that the noise is uncorrelated from pixel to pixel.

\subsection{Evaluation of ${\bf C}^{-1}\cdot{\bf m}$}

The term ${\bf C}^{-1}\cdot{\bf m}$ appears repeatedly in the evaluation of 
(\ref{nr_cl}).  We compute this by solving ${\bf C}\cdot{\bf z} = 
({\bf S}+{\bf N})\cdot{\bf z} = {\bf m}$ for ${\bf z}$.  A more numerically 
tractable system is obtained by multiplying both sides by
${\bf S}^{\frac{1}{2}}\cdot{\bf N}^{-1}$
\begin{equation}
({\bf I} + {\bf S}^{\frac{1}{2}}\cdot{\bf N}^{-1}\cdot{\bf S}^{\frac{1}{2}})
\cdot{\bf S}^{\frac{1}{2}}\cdot{\bf z}
 = {\bf S}^{\frac{1}{2}}\cdot{\bf N}^{-1}\cdot{\bf m}
 = {\bf S}^{\frac{1}{2}}\cdot{\bf y}
\label{czm}
\end{equation}
where ${\bf y}$ is the spherical harmonic transform of the map, defined in 
(\ref{Ny_def}).  Note that ${\bf y}$ can be quickly computed in any 
pixelization scheme, such as HEALPix, that has the property of having pixel 
centers that lie on rings of constant latitude with fixed longitude spacing 
so that fast FFT methods may be used in the  transform \cite{muciaccia}, 
\cite{healpix}.  We then solve (\ref{czm}) using an iterative conjugate 
gradient method with a pre-conditioner for the matrix 
${\bf A} \equiv ({\bf I} + {\bf S}^{\frac{1}{2}}\cdot{\bf N}^{-1}\cdot
{\bf S}^{\frac{1}{2}})$.  We find the following block-diagonal form of ${\bf A}$
to be a good starting point
\begin{equation}
\tilde{{\bf A}} = \left( \begin{array}{cc}
{\bf I} + {\bf S}^{\frac{1}{2}}\cdot\tilde{{\bf N}}^{-1}\cdot{\bf S}^{\frac{1}{2}} & 0 \\
0 & \mbox{diag}({\bf I} + {\bf S}^{\frac{1}{2}}\cdot\tilde{{\bf N}}^{-1}\cdot{\bf S}^{\frac{1}{2}})
\end{array} \right)
\label{pre_con}
\end{equation}
where $\tilde{{\bf N}}^{-1}$ is an approximate block-diagonal form of the noise 
matrix discussed below.  The lower-right block of $\tilde{{\bf A}}$ occupies 
the high $\ell$ portion of the matrix where the signal to noise ratio 
${\bf S}^{\frac{1}{2}}\cdot{\bf N}^{-1}\cdot{\bf S}^{\frac{1}{2}}$ is low, 
so a diagonal approximation is adequate.  The upper-left block occupies the 
low $\ell$ portion of the matrix where the signal dominates the noise, so we 
need a better estimate of ${\bf N}^{-1}$.  In practice we find this split works 
well at $\ell = 512$ for the estimated MAP noise levels.  As for the 
approximate form of ${\bf N}^{-1}$, defined in (\ref{Ny_def}), note that the 
dominant off-diagonal contributions arise from the azimuthally symmetric galaxy 
cut, which couples different $\ell$ modes, but not $m$ modes.  Thus 
${\bf N}^{-1}$ is approximately block diagonal, with perturbations induced by 
the non-uniform sky coverage of MAP.  We therefore use a block diagonal form 
of ${\bf N}^{-1}$ as the pre-conditioner
\begin{equation}
\tilde{N}^{-1}_{(\ell m)(\ell m)^{\prime}} = N^{-1}_{(\ell m)(\ell
m)^{\prime}}\delta_{mm^{\prime}}.
\end{equation}
Using the pre-conditioner (\ref{pre_con}) we find that the conjugate gradient
solution of (\ref{czm}) converges in approximately six iterations and requires
only minutes of processing as a single processor job on an SGI Origins 2000.

\subsection{Evaluation of $\mbox{tr}({\bf C}^{-1}\cdot{\bf P}^{\ell})$ and 
$F_{\ell\ell^{\prime}}$}

There are two approaches to evaluating $\mbox{tr}({\bf C}^{-1}\cdot{\bf P}^{\ell})$.  
The first is to employ the approximate form $\mbox{tr}({\tilde{\bf C}}^{-1}
\cdot{\bf P}^{\ell})$ using the preconditioner (\ref{pre_con}).  The second is
to note that since $\langle {\bf m}\cdot{\bf m}^T \rangle = {\bf C}$, it follows 
that $\mbox{tr}({\bf C}^{-1}\cdot{\bf P}^{\ell}) = \langle {\bf m}^T\cdot
{\bf C}^{-1}\cdot{\bf P}^{\ell}\cdot{\bf C}^{-1}\cdot{\bf m} \rangle$.  Thus we
can use Monte Carlo maps with the requisite signal and noise contributions to
evaluate $\mbox{tr}({\bf C}^{-1}\cdot{\bf P}^{\ell})$.  This is the most 
computationally intensive part of the power spectrum estimation, requiring
$O(N_{mc}N_{iter}N^{\frac{3}{2}})$ operations, where $N_{mc}$ is the number of
Monte Carlo realizations used, and $N_{iter}$ is the number of iterations used
in the conjugate gradient solution of (\ref{czm}).  In practice, we find it
most efficient to use the approximate form for the first few iterations of
(\ref{nr_cl}) then switch to the Monte Carlo form for the final few.

\begin{figure}
\centering
\includegraphics*[width=0.8\hsize,keepaspectratio]{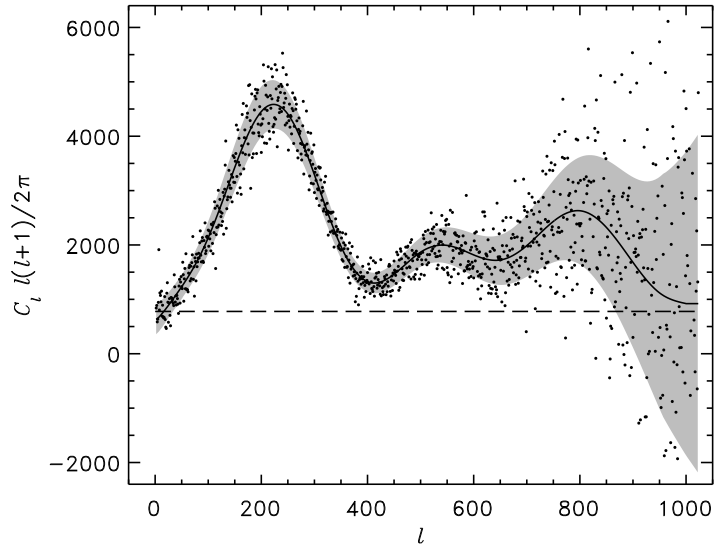}
\caption{\small Sample $C_{\ell}$ spectrum recovered from a simulated 90 GHz 
sky map with MAP-like noise properties.  This spectrum can be shown to be an 
unbiased, nearly minimum variance estimate of the true power spectrum 
\cite{peng}.}
\label{cl_spec}
\end{figure}

The same considerations can be applied to the evaluation of the Fisher matrix, 
however it is important to note that the Newton-Raphson method does not require
an accurate 2nd derivative in order to converge to the correct solution.  Our 
approach to estimating errors in the final spectra are discussed more fully in 
\cite{peng}.  The result of applying this power spectrum estimation to a 
simulated 90 GHz sky map is shown in Fig. \ref{cl_spec}.

\clearpage
\addcontentsline{toc}{section}{Index}
\flushbottom
\printindex

\end{document}